\documentclass[preprint]{article}
\usepackage{neurips_2026}
\usepackage[utf8]{inputenc}
\usepackage[T1]{fontenc}
\usepackage{url}
\usepackage{booktabs}
\usepackage{amsfonts}
\usepackage{amsmath}
\usepackage{graphicx}
\usepackage{xcolor}
\usepackage{algorithm}
\usepackage{algorithmic}
\usepackage{multirow}
\usepackage{subcaption}
\usepackage{enumitem}
\usepackage{tikz}
\usepackage{pgfplots}
\pgfplotsset{compat=1.18}
\usepgfplotslibrary{groupplots}
\usetikzlibrary{arrows.meta, positioning, fit, backgrounds, calc, patterns}

\newcommand{\method}{\mbox{\textsc{SkillAxe}}}

\title{{\method{}}: Sharpening LLM-Authored Agent Skills\\ Through Evaluation-Guided Self-Refinement}

\author{
  Srishti Gautam \\
  Microsoft \\
  \texttt{srgautam@microsoft.com}
  \And
  Arjun Radhakrishna \\
  Microsoft \\
  \texttt{arradha@microsoft.com}
  \And
  Sumit Gulwani \\
  Microsoft \\
  \texttt{sumitg@microsoft.com}
}

\begin{document}

\maketitle

\begin{abstract}
Skill documents, structured natural-language instructions that guide Large Language Model (LLM) agents, are critical to modern agent frameworks, yet LLMs struggle to write skills that actually work. On SkillsBench, human-authored skills improve pass rates by 16.2 percentage points, while LLM-authored skills provide no measurable gain. We introduce \method{}, a fully unsupervised framework that enables LLMs to iteratively diagnose and refine their own skills. \method{} decomposes skill quality into four interpretable dimensions (quality impact, trigger precision, instruction compliance with fault attribution, and solution-path coverage), producing structured improvement briefs that require no ground-truth labels, test suites, or environment rewards. On SkillsBench, \method{} improves pass rates by 28\% relative over unimproved LLM skills and closes 47--67\% of the gap to human-authored skills. We validate the approach as a continuous improvement engine in the wild on SpreadsheetBench, where a \method{}-built skill library learns from past agent trajectories and raises pass rate from 16.0\% to 52.0\% using only 22 skills.
\end{abstract}

\section{Introduction}

LLM-powered agents increasingly rely on \emph{skills}, structured natural-language instructions that encode domain knowledge, API usage patterns, and procedural strategies~\cite{anthropicskills2025,agentskillssurvey2026,sokskills2026}. Skills are a primary lever for improving agent performance: on SkillsBench~\cite{skillsbench2026}, human-authored skills increase pass rates by 16.2 percentage points. Yet skill development today remains largely \emph{manual}, \emph{anecdotal}, and \emph{opaque}~\cite{anthropicskills2025,openaiskills2026,lcskills2026}. Practitioners iteratively tweak prompts and inspect outputs qualitatively, often without knowing whether failures stem from flawed instructions, incorrect triggering, or agents ignoring useful guidance. This lack of diagnostic feedback makes systematic skill refinement difficult not only for automated pipelines, but also for non-expert human authors.

This limitation becomes especially visible for LLM-authored skills, where refinement must occur without expert oversight. SkillsBench shows that LLM-authored skills provide \emph{no measurable improvement} over bare agents, despite being syntactically fluent and often superficially plausible. Existing evaluations provide only coarse task-level pass/fail signals~\cite{skillsbench2026,skillswild2026,skillcraft2026}, even though skill failures are often systematic: triggers activate on the wrong tasks, instructions conflict with constraints, or plausible guidance silently leads agents astray. Without actionable diagnostics, both human and automated skill refinement remain brittle trial-and-error processes.

These limitations become especially consequential in procedural domains where agents must execute long, tool-dependent workflows rather than produce short-form text outputs. Spreadsheet manipulation is particularly challenging because agents must coordinate APIs, formulas, recalculation semantics, and multi-step execution pipelines. We study this setting using Excel Copilot~\cite{ms}, a production-grade agent with multi-turn reasoning and dynamic tool selection. Despite this sophistication, failures still frequently stem from missing procedural knowledge: writing formulas without triggering recalculation, inserting VBA code as spreadsheet text rather than executable macros, or selecting incorrect workbook manipulation strategies. SpreadsheetBench~\cite{spreadsheetbench2024} captures these challenges through 912 real-world tasks. Our experiments show that even highly capable agents with 10+ tool calls per task can discover and follow reusable skill guidance when skills are exposed as optional, on-demand documents. However, existing approaches provide little insight into which skills help, why they fail, or how they should be systematically improved once deployed.

\begin{figure}[t]
\centering
\includegraphics[width=\textwidth]{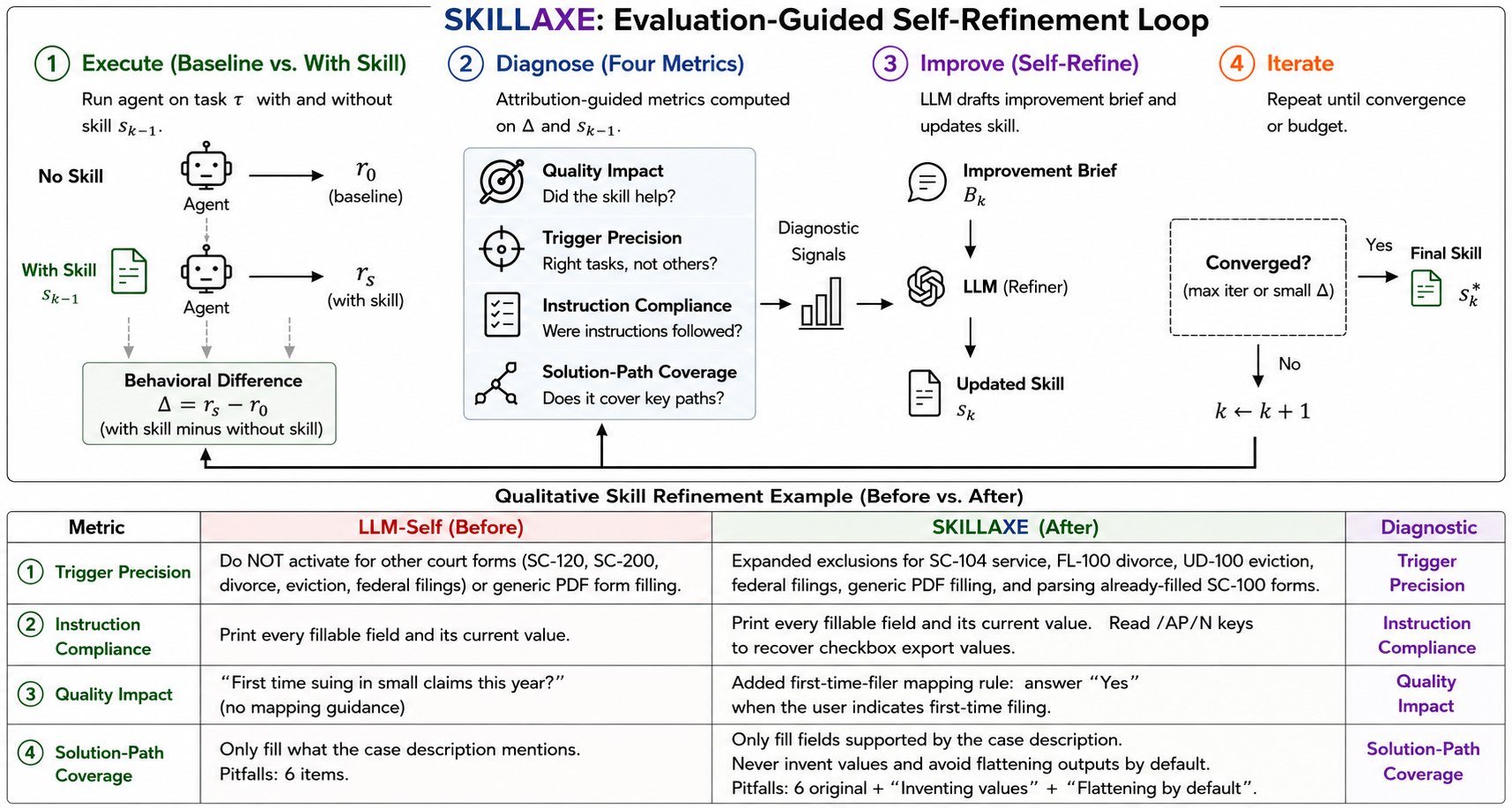}
\caption{\textbf{[Top]}~\method{} overview. The agent runs each task with and without the current skill (Phase~1). Four unsupervised metrics diagnose quality impact, trigger precision, instruction compliance, and solution-path coverage (Phase~2). An LLM refiner uses the resulting improvement brief to produce an updated skill (Phase~3), iterating until convergence (Phase~4). \textbf{[Bottom]}~Qualitative example: \textbf{court-form-filling} (reward: 0.50\,$\rightarrow$\,1.00). Each row shows a before/after change traceable to a specific metric.}
\label{fig:framework}
\end{figure}

We address this with \method{} (Figure~\ref{fig:framework}), an evaluation-guided self-refinement framework for agent skills. \method{} runs the agent on the same task with and without the current skill, diagnoses the resulting behavioral differences across four complementary dimensions: (i)~\emph{quality impact}, whether the skill improves outcomes; (ii)~\emph{trigger precision}, whether it activates for the right tasks; (iii)~\emph{instruction compliance}, whether the agent follows the skill's guidance; and (iv)~\emph{solution-path coverage}, whether the skill supports the range of plausible execution strategies available to the agent. These diagnostics are compiled into a structured improvement brief used to rewrite the skill. The key insight is that many skill documents implicitly specify evaluable behavioral constraints: by treating the skill's rules as evaluation criteria and contrasting outputs with and without the skill, \method{} produces actionable feedback without requiring task-specific labels, handcrafted reward functions, or human-written test suites.

On SkillsBench, \method{} raises pass rate by 28\% relative over unimproved LLM-authored skills (41.6\% vs.\ 32.5\%), closing 47--67\% of the gap to human-authored skills. The improvement is driven primarily by \emph{execution reliability}: coverage rises from 49.4\% to 72.7\% while correctness among completed tasks remains comparable. We further evaluate \method{} on SpreadsheetBench in an in-the-wild continual-learning setting, where Excel Copilot continuously accumulates reusable skills from prior task trajectories. Starting from no curated library, \method{} iteratively refines spreadsheet skills over time, producing a compact 22-skill library that raises pass rate from 16.0\% to 52.0\% on held-out tasks.

\paragraph{Contributions.}
\begin{itemize}[leftmargin=*]

\item We identify that the central limitation of current skill engineering is the lack of actionable diagnostic feedback. Through analysis on SkillsBench, we show that LLM-authored skills fail in systematic ways not captured by scalar task-success signals, including conflicting instructions and missing execution knowledge.

\item We introduce \method{}, an evaluation-guided framework for refining agent skills through diagnostic feedback derived from agent behavior. The framework isolates distinct failure modes including incorrect triggering, weak instructions, agent non-compliance, and narrow solution support without requiring task-specific supervision.

\item We empirically validate \method{} on SkillsBench and SpreadsheetBench, showing substantial improvements over unimproved LLM-authored skills in both controlled evaluation and continual in-the-wild skill accumulation settings.

\end{itemize}

\section{Related Work}
\label{sec:related}

\textbf{Skill libraries and ecosystems.}
Recent work has explored reusable skill libraries for embodied and interactive agents~\cite{voyager2024,sage2025,skillact2024}, alongside infrastructures for retrieving, and evolving skills over time~\cite{agentskillssurvey2026,sokskills2026,skillrouter2026,skillnet2026,cosplay2026}. While these systems show that reusable skills substantially improve agent performance, most focus on storing or selecting skills rather than refining their internal structure. Approaches such as Voyager~\cite{voyager2024} and SAGE~\cite{sage2025} incorporate iterative improvement, but rely on environment-specific reward signals unavailable for natural-language skills in open-ended domains.
\\
\textbf{Skill evaluation, optimization, and self-improvement.}
Existing skill benchmarks evaluate skills primarily through task-level success signals~\cite{skillsbench2026,skillswild2026,skillcraft2026}, providing limited insight into why a skill failed or whether the failure originated from the skill itself. Other approaches optimize token efficiency~\cite{skillreducer2026}, skill selection~\cite{skillmoo2026}, or prompts using supervised reward signals~\cite{zhou2023ape,yang2024opro,alignpro}. TextGrad~\cite{yuksekgonul2024textgrad} and Trace~\cite{cheng2024trace} generate richer gradient-like textual feedback, while Reflexion~\cite{reflexion2023}, Self-Refine~\cite{selfrefine2023}, and AgentRefine~\cite{fu2025agentrefine} improve transient executions or model behavior rather than persistent skill documents. In contrast, \method{} refines reusable skills through diagnostic feedback derived directly from agent behavior without requiring task-specific supervision. \\
\textbf{Failure attribution.}
Prior work studies failure attribution across interacting agents~\cite{zhang2025failureattribution} and instruction-following verification~\cite{lou-etal-2024-large}. \method{} instead attributes failures at the boundary between the skill and the agent, distinguishing flawed skill instructions from agent execution errors to support targeted rewriting while preserving correct rules.

\section{\method{}: Diagnostic Skill Evaluation and Self-Refinement}
\label{sec:framework}

We evaluate a skill $s$ for a task $\tau$ with instruction $q$ by comparing the agent's response without the skill, $r_0$, against the response with the skill injected, $r_s$. \method{} computes four complementary diagnostics from $(q, r_0, r_s, s)$ without requiring task-specific labels, handcrafted reward functions, or expert-written grading rubrics. LLMs are used as general-purpose judges and rewriters rather than supervised evaluators tied to a particular environment or benchmark. Each diagnostic isolates a different failure boundary in the skill lifecycle: whether the skill helps at all, whether it activates correctly, whether the agent follows it, and whether the skill captures sufficient execution knowledge.

\subsection{Quality Impact}
\label{sec:quality}

Before diagnosing \emph{why} a skill failed, we must first determine whether the skill helped at all. A skill may appear fluent, detailed, and topically relevant yet still degrade performance by steering the agent toward an incorrect solution strategy. Quality Impact therefore serves as the outer-loop evaluation signal: it measures the overall behavioral effect of the skill before the remaining diagnostics localize the cause. We evaluate the marginal contribution of the skill through paired comparison between the agent's response with the skill, $r_s$, and the baseline response without the skill, $r_0$, in two stages.\\
\textbf{Stage 1: Preference direction.}
An LLM judge receives both responses in randomized order and determines which better satisfies the task instruction $q$, producing:
$
d \in \{r_s \succ r_0,\; r_0 \succ r_s,\; \text{tie}\}.
$\\
\textbf{Stage 2: Improvement magnitude.}
Conditioned on the preferred direction, the judge estimates the strength of the improvement:
$
\text{Quality}(s,\tau)
=
d \cdot m
\in [-1,1],
$
where $d \in \{-1,0,+1\}$ encodes direction and $m \in [0,1]$ estimates magnitude.

The decomposition into direction and magnitude is important for refinement. A small regression suggests local rewriting, while a catastrophic failure indicates that the skill may contain fundamentally incorrect guidance or severe trigger errors. For example, a chart-generation skill may consistently produce visually polished outputs while selecting the wrong chart type. In this case the skill receives a negative Quality Impact score even though parts of the skill remain useful, signaling that the refinement process should revise chart-selection logic rather than discard the entire skill. The remaining diagnostics, i.e, Trigger Precision, Instruction Compliance, and Solution-Path Coverage, are then used to determine the source of the regression.
\begin{figure}[t]
\centering
\includegraphics[width=\textwidth]{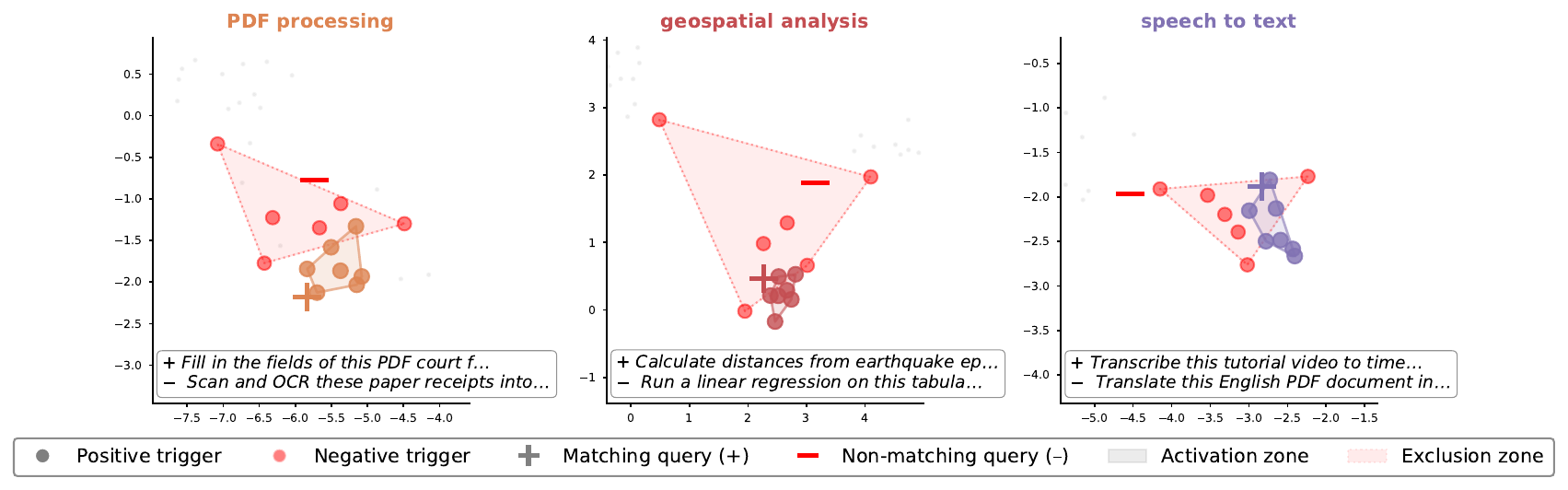}
\caption{UMAP projection of trigger embeddings for three SkillsBench skills. Positive trigger phrases form activation regions (solid hulls); exclusions form neighboring rejection regions (dotted hulls). Task queries near the decision boundary expose ambiguous or over-broad triggers.}
\label{fig:trigger_embedding}
\end{figure}

\subsection{Trigger Precision}
\label{sec:trigger}

A common source of negative quality impact is incorrect activation: the skill fires on tasks outside its intended scope. Modern agent frameworks already perform skill routing through semantic matching over natural-language skill descriptions~\cite{anthropicskills2025,skillrouter2026}, often using YAML trigger fields or LLM-based retrieval. \method{} evaluates trigger quality in the same semantic retrieval space.

We first extract positive trigger phrases $P^+$ (when the skill should activate) and negative trigger phrases $P^-$ (when it should not) using dependency parsing with negation tracking. Each phrase is then represented as a \emph{context-aware vector} formed by blending:
(i)~a phrase embedding capturing the semantic meaning of the phrase itself, and
(ii)~a sentence-context embedding capturing how the phrase is used inside the skill description, i.e, the logic.

The contextual component is critical because the same phrase may appear in both activation and exclusion contexts. For example, ``spreadsheet'' in ``create a new spreadsheet from scratch'' should lie near the activation region, while ``Do NOT trigger for chart images without spreadsheets'' should lie near the exclusion boundary. Without sentence context these phrases remain artificially similar; contextual embeddings separate them geometrically and produce clearer retrieval boundaries. This follows the broader principle used in self-explainable representation learning: explanations should be evaluated in the same semantic space used for downstream decisions~\cite{NEURIPS2022_722f3f92}. The resulting embedding space defines an \emph{activation region} for valid tasks and an \emph{exclusion region} for invalid ones (Figure~\ref{fig:trigger_embedding}). We compute three geometric diagnostics corresponding to three distinct failure modes:
\\[0.5em]
\textbf{Coverage breadth.}
A trigger may be too narrow and only activate for a small subset of relevant tasks. We measure the mean pairwise cosine distance between all positive trigger embeddings:
$\text{CoverageBreadth} = \frac{1}{|P^+|(|P^+|-1)} \sum_{i \neq j} \bigl(1-\cos(\mathbf{e}_i^+,\mathbf{e}_j^+)\bigr)$,
where $\mathbf{e}_i^+$ is the embedding of the $i$-th positive trigger phrase.
\\[0.5em]
\textbf{Negative specificity.}
A trigger may also activate for neighboring but incompatible tasks. For each negative trigger phrase, we compute the distance to its nearest positive trigger:
$\text{NegativeSpecificity} = \frac{1}{|P^-|} \sum_{\mathbf{e}^- \in P^-} \min_{\mathbf{e}^+ \in P^+} \bigl(1-\cos(\mathbf{e}^-,\mathbf{e}^+)\bigr)$.
Low specificity indicates that exclusion phrases remain too close to the activation region.
\\[0.5em]
\textbf{Boundary sharpness.}
Even if average separation is good, a single ambiguous trigger can cause catastrophic misrouting. We measure the minimum positive-negative distance:
$\text{BoundarySharpness} = \min_{\mathbf{e}^- \in P^-, \mathbf{e}^+ \in P^+} \bigl(1-\cos(\mathbf{e}^-,\mathbf{e}^+)\bigr)$.
Low sharpness exposes worst-case ambiguity at the decision boundary.

Together, these metrics distinguish narrow triggers, weak exclusions, and fuzzy boundaries.

\begin{table}[t]
\centering
\caption{Fault attribution in action. The same observed behavior (wrong shade of yellow) leads to different improvement actions depending on rule specificity.}
\label{fig:fault_example}
\footnotesize
\begin{tabular}{@{}lll@{}}
\toprule
& \textbf{Specific rule $\rightarrow$ agent fault} & \textbf{Vague rule $\rightarrow$ skill fault} \\
\midrule
\textbf{Rule} & ``Use yellow background (\texttt{FFFF00}) for assumptions'' & ``Use yellow background for assumptions'' \\
\textbf{Evidence} & Agent used \texttt{FFF2CC} instead of \texttt{FFFF00} & Agent used \texttt{FFF2CC} (a valid yellow) \\
\textbf{Quality} $g_i$ & 1.0 & 0.2 \\
\textbf{Adherence} $a_i$ & 0.3 & 0.7 \\
\textbf{Skill-fault} $f_i$ & 0.0 & 0.3 \\
\textbf{Skill credit} $c_i$ & $0.3 + 0.7 \times 1.0 = 1.0$ & $0.7 + 0.3 \times 0.7 = 0.91$ \\
\midrule
\textbf{Action} & Preserve rule (agent execution error) & Sharpen rule (add hex code) \\
\bottomrule
\end{tabular}
\end{table}

\subsection{Instruction Compliance with Fault Attribution}
\label{sec:compliance}

Even when a skill activates on the correct task, failures can arise for two fundamentally different reasons: the skill itself may provide poor guidance, or the agent may simply fail to follow otherwise correct instructions. Existing evaluation approaches typically collapse these cases into a single failure signal, implicitly treating every bad outcome as evidence that the skill should be rewritten~\cite{}. In practice, this can degrade useful skills by ``fixing'' rules that were already correct.

\method{} therefore separates \emph{instruction adherence} from \emph{skill quality}. The central question is not only whether the agent followed the skill, but whether the skill deserved to be followed in the first place. We begin by decomposing the skill into explicit evaluable rules, following the general idea of rubric-based evaluation, but with the rubric extracted directly from the skill document itself rather than provided externally. Given only the skill $s$, an LLM extracts procedural rules $\{R_1,\ldots,R_n\}$ together with severity weights $w_i \in \{1,2,3\}$ indicating whether violations correspond to minor, major, or critical constraints.

Each rule is then evaluated against the task instruction $q$ and the agent output $r_s$. For every rule $R_i$, the evaluator predicts:
\vspace{-0.3cm}
\begin{itemize}[leftmargin=*]
    \item \textbf{Relevance} $\rho_i \in \{0,1\}$: whether the rule applies to the current task.\vspace{-0.15cm}
    \item \textbf{Adherence} $a_i \in [0,1]$: how well the agent followed the rule.\vspace{-0.15cm}
    \item \textbf{Rule quality} $g_i \in [0,1]$: whether the rule itself is precise and operationalizable.\vspace{-0.15cm}
    \item \textbf{Skill fault} $f_i \in [0,1]$: whether observed failures originate from weak skill guidance or from the agent ignoring otherwise valid instructions.
\end{itemize}

Evaluation is grounded in file-level evidence such as formulas, formatting codes, and structural diffs rather than the agent's textual claims about its behavior. Figure~\ref{fig:fault_example} illustrates the central distinction: violating a precise rule specifying an exact RGB code indicates agent failure, whereas violating a vague rule like ``use yellow'' suggests that the skill itself lacks sufficient specificity.

We first compute raw instruction adherence:
$
\text{Compliance}(s,\tau)
=
\frac{
\sum_{i:\rho_i=1} w_i g_i a_i
}{
\sum_{i:\rho_i=1} w_i
},
$
where relevance $\rho_i$ filters non-applicable rules, adherence $a_i$ measures execution fidelity, rule quality $g_i$ downweights vague instructions, and severity weights $w_i$ prioritize critical constraints.

However, low adherence alone does not necessarily imply a poor skill. A well-written rule may still be ignored by the agent. To separate these cases, we define a fault-adjusted skill credit:
$
c_i
=
a_i + (1-a_i)(1-f_i),
$
where $f_i$ estimates whether failures are attributable to the skill itself. When failures are judged to originate from the agent rather than the skill, credit is partially restored.

Replacing adherence with fault-adjusted credit yields the final skill-quality score:\vspace{-0.2cm}
\[
\text{SkillScore}(s,\tau)
=
\frac{
\sum_{i:\rho_i=1} w_i g_i c_i
}{
\sum_{i:\rho_i=1} w_i
}.
\]
This decomposition is critical for refinement. A low Compliance score indicates that the agent failed to follow instructions, while a low SkillScore indicates that the instructions themselves require rewriting. The distinction prevents \method{} from degrading correct rules simply because the agent executed them poorly.

\subsection{Solution-Path Coverage}
\label{sec:solution_paths}

The previous diagnostics evaluate whether a skill helps, activates correctly, and provides followable instructions. However, even a correct and well-executed skill may still be ineffective if it only supports a narrow subset of valid solution strategies.

Agent tasks often admit multiple execution paths. A spreadsheet task, for example, may be solved through:
(i)~dataframe-centric processing with pandas,
(ii)~direct workbook manipulation with \texttt{openpyxl},
(iii)~Excel COM automation, or
(iv)~formula-driven editing within the spreadsheet itself. Skills are not executable programs that deterministically prescribe a single trajectory; instead, they act as guidance priors over a broader space of possible agent behaviors. If the agent pursues a valid strategy that the skill does not meaningfully support, the skill becomes operationally useless even when activated correctly. \method{} therefore evaluates whether the skill covers the range of plausible solution strategies for the task rather than overcommitting to a single procedural path.

Given the task instruction $q$ and the skill type, an LLM first enumerates plausible solution paths
$\{P_1,\ldots,P_k\}$.
We then measure how well chunks of the skill align with each path:
$\text{PathCoverage}(s,\tau) = \frac{1}{k} \sum_{j=1}^{k} \max_{c \in \mathrm{chunks}(s)} \cos(\mathbf{e}(P_j),\mathbf{e}(c))$,
where $\mathbf{e}(P_j)$ is the embedding of the $j$-th solution-path description and $\mathbf{e}(c)$ is the embedding of a skill chunk.

The score measures whether each plausible execution strategy is represented somewhere in the skill guidance. Low coverage indicates that the skill only supports a narrow behavioral manifold, leaving the agent without useful guidance when it explores alternative but valid solution paths.

This diagnostic complements the earlier metrics: a skill may trigger correctly and contain precise instructions, yet still fail systematically because its guidance only aligns with a limited subset of the agent's possible execution strategies.

\section{Experiments}
\label{sec:experiments}

We evaluate \method{} along two axes that correspond to distinct deployment scenarios for skill improvement:

\begin{enumerate}[leftmargin=*]
    \item \textbf{Task-specific improvement} (Section~\ref{sec:setup}--\ref{sec:fair_grading}): Can \method{} improve a skill written for a known task, where the task instruction is available during evaluation? We test this on SkillsBench~\cite{skillsbench2026} (77 tasks, 191 skills, 688 agent runs), comparing \method{}-improved skills against unimproved LLM skills and human-authored references.
    \item \textbf{Continuous library improvement} (Section~\ref{sec:transfer}): Can \method{} serve as a flywheel for production agents, learning from past trajectories to build an ever-improving skill library? We test this on SpreadsheetBench~\cite{spreadsheetbench2024} (200 train / 50 test split), where \method{} builds a skill library from training trajectories and routes skills to unseen test tasks.
\end{enumerate}


\subsection{SkillsBench Setup}
\label{sec:setup}

\paragraph{Benchmark.} SkillsBench~\cite{skillsbench2026} contains 89 tasks spanning software engineering, data science, scientific computing, and domain-specific applications. Each task has 1--7 skill slots (median: 2), with 62 tasks using multiple co-deployed skills, totaling 191 unique skills. Tasks are executed in Docker containers via the Harbor runner with Claude Opus 4.5 as the agent backbone. Six tasks requiring external API credentials are excluded, leaving 83 eligible tasks, of which 77 yielded at least one successful run across conditions. Each condition is run $k{=}2$ times per task, following SkillsBench's Terminal-Bench Method~D scoring~\cite{skillsbench2026}: per-task mean reward across $k$ trials, then mean across all 77~tasks with a fixed denominator. 

\paragraph{Conditions.} Following the SkillsBench protocol, skills are \emph{task-specific}: each skill slot receives a dedicated skill document matched to that slot's domain. We compare four conditions: \textbf{No-skill}: Bare agent with no skill documents, \textbf{LLM-self}: LLM-authored skills generated by the SkillsBench prompt~\cite{skillsbench2026}, one per skill slot, \textbf{\method{}}: LLM-self skills improved by one iteration of the \method{} loop. Each of the 191 skills is evaluated and improved independently using its task as the evaluation context and \textbf{Human}: Human-authored reference skills from SkillsBench.

\paragraph{Evaluation.} We report results under two evaluation protocols: (1)~a \emph{multimodal fair grader} that evaluates agent outputs without access to human-written test suites, using an LLM judge~\cite{liu2023geval,zheng2023mtbench}, and (2)~the native SkillsBench verifiers as a conservative lower bound. This dual evaluation lets us measure true task completion while maintaining comparability with prior work.

\subsection{Multimodal Fair Grading}
\label{sec:fair_grading}

SkillsBench's native verifiers are human-authored pytest suites. While consistent, these tests encode assumptions beyond actual task requirements: they assert that Excel formulas trigger recalculation, that cell values match to exact decimal precision, or that specific metadata fields are populated. When an agent solves a task correctly via a different approach, the native verifier rejects it.

\paragraph{Protocol.} For each of the runs, the fair grader receives: (1) the task instruction, (2) input files the agent started with, and (3) the agent's actual output files, rendered as images where possible (Excel sheets, PDFs, charts) or as text. It does \emph{not} see the verifier test suite, the skill document, the native reward, or the agent's trajectory. The grader requires \emph{concrete output artifacts} that substantively satisfy the task. Minor formatting differences ($\sim$1\%) are tolerated, but plans, narration, or code without results are insufficient. Full protocol details are in Appendix~\ref{app:fairgrader}.

\begin{figure}[t]
\centering
\includegraphics[width=0.8\textwidth]{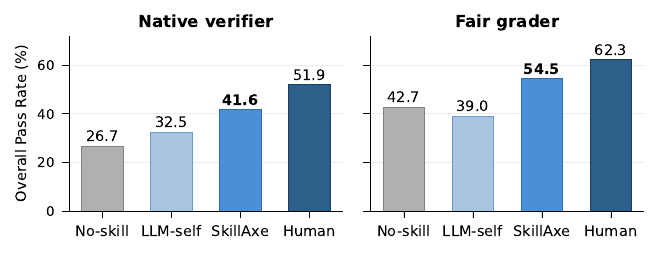}
\caption{Overall pass rates on SkillsBench (77 tasks, $k{=}2$ trials, Method~D). Under both evaluation protocols, \method{} substantially outperforms unimproved LLM skills and closes 47--67\% of the gap to human-authored skills.}
\label{fig:main_results}
\end{figure}

\paragraph{Results.} Figure~\ref{fig:main_results} presents the main results using Terminal-Bench Method~D~\cite{skillsbench2026} scoring across 77~tasks ($k{=}2$ trials per task). We decompose overall pass rate hierarchically into \emph{coverage} (fraction of tasks where the agent produces evaluable output) and \emph{quality} (pass rate among tasks that produce output), since overall $=$ coverage $\times$ quality. Table~\ref{tab:hierarchical} provides the full decomposition.

\begin{table}[t]
\centering
\caption{Hierarchical decomposition of pass rates (77 tasks, $k{=}2$ trials, Method~D). \emph{Coverage}: fraction of tasks producing evaluable output. \emph{Quality}: pass rate among output-producing tasks. \emph{Overall} $=$ Coverage $\times$ Quality. The entire \method{} gain over no-skill comes from coverage (+26pp), not quality (both 57.1\%).}
\label{tab:hierarchical}
\begin{tabular}{lccccc}
\toprule
\textbf{Condition} & \textbf{Coverage} & \textbf{Native Quality} & \textbf{Fair Quality} & \textbf{Cov$\times$Nat} & \textbf{Cov$\times$Fair} \\
\midrule
No-skill       & 46.7\% & 57.1\% & 91.4\% & 26.7\% & 42.7\% \\
LLM-self       & 49.4\% & 65.8\% & 78.9\% & 32.5\% & 39.0\% \\
\method{}      & \textbf{72.7\%} & 57.1\% & 75.0\% & \textbf{41.6\%} & \textbf{54.5\%} \\
Human          & 68.8\% & \textbf{75.5\%} & \textbf{90.6\%} & 51.9\% & 62.3\% \\
\bottomrule
\end{tabular}
\end{table}

Three findings emerge. First, \method{}-improved skills dramatically increase agent coverage: 72.7\% of tasks produce evaluable output vs.\ 46.7\% for no-skill. This means agents reliably complete hard tasks that crash or time out without skills, producing concrete artifacts on 56 vs.\ 36 tasks.

Second, a hierarchical decomposition reveals a precise mechanism: \emph{among tasks where agents produce output}, native quality is identical for no-skill and \method{} (both 57.1\%). The entire overall gain (41.6\% vs.\ 26.7\%) comes from coverage, not answer quality. Skills contain \emph{execution knowledge}, procedural guidance that helps agents find the right libraries, handle setup, and avoid crashes, rather than improving the quality of answers the agent already knows how to produce.

Third, under the fair grader, \method{} achieves the highest overall rate (54.5\%), with fair quality at 75.0\% among output-producing tasks. Human skills achieve the highest quality on both metrics (75.5\% native, 90.6\% fair), confirming that human expertise provides both execution knowledge and answer-quality guidance.

These results are corroborated under SkillsBench's native verifiers with Method~D scoring (mean continuous reward): \method{} achieves 27.3\% vs.\ 24.9\% no-skill and 25.8\% LLM-self, closing 11.4\% of the gap to human skills (46.0\%). Full Method~D results appear in Appendix~\ref{app:method_d}. Figure~\ref{fig:framework} (bottom) presents a qualitative before/after refinement example illustrating how each diagnostic contributes to transforming an initially weak skill into a more precise, executable, and robust one.

\subsection{Transfer to SpreadsheetBench: Skill Libraries in the Wild}
\label{sec:transfer}

SkillsBench evaluates task-specific skills assigned per task--a controlled setting that isolates skill quality. In production, however, the value proposition is different: a complex agent like Excel Copilot serves thousands of diverse requests daily, and maintaining a bespoke skill per task is infeasible. Instead, the agent draws from a shared \emph{skill library} that grows and improves continuously as the system encounters new failure patterns. This is the deployment scenario \method{} is designed for: past agent trajectories reveal what went wrong, \method{}'s diagnostics identify \emph{which part} of which skill caused the failure, and the improvement loop sharpens the library without human curation--a flywheel where every failed run makes the next one more likely to succeed.

We evaluate this setting on SpreadsheetBench~\cite{spreadsheetbench2024}, a benchmark of 912 real-world spreadsheet manipulation tasks with cell-level golden answers. Spreadsheet tasks are an ideal testbed: they span a long tail of operations (formula writing, conditional formatting, VBA macros, pivot tables) where no single prompt can anticipate every failure mode, making a continuously improving skill library essential. We split SpreadsheetBench into 200 train / 50 test tasks (random, seed=42). The agent is an Excel copilot powered by Claude Opus 4.5; skill library construction (generation, routing, and improvement) uses Claude Opus 4.5 as the backbone LLM.

\paragraph{Skill library construction.} We compare three conditions, similar to SkillsBench: (1)~\textbf{No skill}: the bare Excel copilot agent without any skill injection; (2)~\textbf{LLM-self library}: 69 skills built from the same generation prompt without improvement---for each training task, an LLM call decides whether to create a new skill or skip (binary fork/skip), producing one narrow skill per task pattern; (3)~\textbf{\method{} library}: 22 skills built with embedding-based 3-zone routing (FORK/IMPROVE/SKIP) that uses positive and negative trigger phrases to decide whether each new task should create a new skill, improve an existing one, or be skipped---then two-phase improvement using trigger diagnostics and solution-path coverage (Phase~1) followed by quality impact and instruction compliance from agent trajectories (Phase~2). The \method{} routing mechanism naturally clusters tasks by embedding similarity, merging related patterns into broader skills rather than creating one skill per task. Improvements are capped at 5 training tasks per skill to prevent overfitting to narrow patterns; Appendix~\ref{app:ssb_routing} shows the resulting distribution. Both libraries are built from the 200 train tasks only; all results are reported on the held-out test tasks.

\paragraph{Skill activation.} At test time, skills are \emph{discoverable but optional}: the agent sees a manifest of available skill names and descriptions, and may call a \texttt{get\_skill\_md} tool to load any skill before planning its approach. This mirrors production deployment where the agent autonomously decides whether a skill is relevant. The \method{} library exposes 22 skills; the LLM-self library exposes all 69.

\paragraph{Results.} Table~\ref{tab:ssb_results} summarizes results on the held-out test tasks evaluated on a common subset where all three conditions completed ($N=50$). Evaluation uses SpreadsheetBench's native verifier, which performs cell-level comparison of agent output against golden-answer workbooks.

\begin{table}[h]
\centering
\caption{SpreadsheetBench results on common test tasks ($N=50$). An Excel Copilot agent (Claude Opus 4.5 as backbone) is evaluated under three conditions. Both skill libraries dramatically improve over the no-skill baseline; \method{} matches LLM-self accuracy with 68\% fewer skills.}
\label{tab:ssb_results}
\begin{tabular}{lccc}
\toprule
\textbf{Condition} & \textbf{Pass Rate} & \textbf{Skills} & \textbf{Activation} \\
\midrule
No skill                          & 16.0\% & --- & --- \\
LLM-self library (69 skills)      & 52.0\% & 69 & 20.0\% \\
\method{} library (22 skills)     & 52.0\% & 22 & 35.8\% \\
\bottomrule
\end{tabular}
\end{table}

Both skill libraries improve dramatically over the no-skill baseline (+36.0pp, from 16.0\% to 52.0\% - \ref{tab:ssb_results}), confirming that skill injection is transformative for complex spreadsheet tasks. The \method{} library matches the naive library's accuracy while using 68\% fewer skills (22 vs.\ 69). Crucially, \method{} skills are activated nearly twice as often (35.8\% vs.\ 20.0\%), indicating that the consolidated, improved descriptions are more recognizable to the agent. This efficiency gain matters in production: fewer skills reduce the manifest size injected into every prompt, lowering token cost and latency, while higher activation means the skills that exist are actually used. Per-query inspection confirms that both libraries pass and fail on the same tasks despite producing different workbooks, suggesting the accuracy ceiling is determined by task difficulty rather than skill content---the primary benefit of \method{} is compression, not accuracy. Intrinsic trigger analysis confirms that \method{} descriptions produce 3$\times$ wider discrimination margins than human skills (Appendix~\ref{app:trigger_analysis}).

\section{Discussion}
\label{sec:discussion}

\paragraph{Skills as execution knowledge.}
The hierarchical decomposition (Table~\ref{tab:hierarchical}) shows that skills primarily improve \emph{execution reliability}, not answer quality. Native correctness among completed tasks remains unchanged with and without \method{} skills (57.1\%), while coverage rises from 46.7\% to 72.7\%. Skills encode procedural knowledge---library usage, format handling, recalculation semantics, and workflow structure---that prevents brittle execution failures (Appendix~\ref{app:exec_example}).

\paragraph{Continuous deployment-time refinement.}
Because the diagnostics require only the task instruction and agent outputs, \method{} can evaluate and refine skills wherever agents are deployed, without handcrafted test suites or environment-specific reward functions. The SpreadsheetBench experiments demonstrate this through continual accumulation and refinement of reusable spreadsheet skills over time.

\paragraph{Test-suite bias.}
The divergence between fair and native evaluation (Table~\ref{tab:hierarchical}) highlights a limitation of rigid verifier-based benchmarks. Under fair grading, \method{} achieves 54.5\% versus 42.7\% for the no-skill baseline, suggesting that improved skills often guide agents toward valid alternative solution strategies rejected by task-specific test suites.

\section{Limitations}
\label{sec:limitations}

\method{} has several limitations. First, rule-level refinement cannot detect \emph{structural misalignment}: a skill may teach a fundamentally incorrect strategy while remaining internally self-consistent, producing few detectable rule violations. Second, skills are currently evaluated independently, and interactions or conflicts between multiple simultaneously active skills are not explicitly modeled. Third, both compliance evaluation and fair grading rely on LLM judges, which may exhibit positional or calibration biases~\cite{zheng2023mtbench}. Finally, although \method{} substantially improves over unimproved LLM-authored skills, the remaining gap to human-authored skills likely requires stronger online execution feedback and multi-iteration refinement beyond the current single-update loop.
\section{Conclusion}
\label{sec:conclusion}

We presented \method{}, a fully unsupervised framework that decomposes skill quality into four interpretable dimensions and produces structured improvement briefs an LLM can act on. On SkillsBench, \method{} raises pass rate by 28\% relative and closes 47--67\% of the gap to human skills, primarily by improving execution reliability. Transfer to SpreadsheetBench validates the approach as a continuous improvement engine for production skill libraries. \method{} reduces the need for human expertise in skill authoring: any LLM can produce a draft skill, and the diagnostic loop automatically identifies what is wrong and how to fix it.

\bibliographystyle{unsrt}
\bibliography{references}

\newpage
\appendix

\section{Broader Impacts}

This work aims to improve the reliability, diagnosability, and maintainability of LLM-powered agents by enabling systematic refinement of reusable skill documents. More reliable skills can reduce brittle agent behavior in procedural domains such as spreadsheet manipulation, document processing, and tool-use workflows, where failures often arise from missing execution knowledge or ambiguous instructions. By providing localized diagnostic feedback, \method{} may also help non-expert users iteratively improve agent behavior without requiring extensive prompt-engineering expertise.

At the same time, automatically refined skills may propagate incorrect procedural guidance if the underlying evaluations are themselves imperfect. Since \method{} relies on LLM-based judges and heuristic diagnostics, incorrectly attributed failures or flawed refinement suggestions could reinforce undesirable behaviors over repeated optimization cycles. In high-stakes domains, over-reliance on automatically generated skills without human verification may therefore introduce reliability or safety risks.

The framework is intended as a diagnostic and refinement tool rather than a replacement for human oversight. We recommend that automatically refined skills, particularly those deployed in production or safety-critical settings, be reviewed and validated by developers before deployment.

\section{Per-Task Results}
\label{app:pertask}

Table~\ref{tab:winners} lists the SkillsBench tasks with the largest improvement from LLM-self to \method{} under the native verifier (Method~D scoring). These represent cases where \method{}'s improvement loop added critical procedural guidance that was absent from the initial LLM-authored skill. Notably, 5 of the 8 tasks improve from zero reward, indicating that the original skill provided no useful guidance and \method{}'s diagnostics identified actionable fixes. The largest gain is \emph{court-form-filling} (+0.50), where the compliance metric identified that the agent was ignoring field-ordering constraints specified in the skill.

\begin{table}[h]
\centering
\caption{Tasks with largest improvement from LLM-self to \method{} (native verifier, matched set).}
\label{tab:winners}
\begin{tabular}{lccc}
\toprule
\textbf{Task} & \textbf{LLM-self} & \textbf{\method{}} & \textbf{$\Delta$} \\
\midrule
court-form-filling & 0.50 & 1.00 & +0.50 \\
mario-coin-counting & 0.00 & 0.33 & +0.33 \\
lake-warming-attribution & 0.00 & 0.33 & +0.33 \\
invoice-fraud-detection & 0.00 & 0.29 & +0.29 \\
pptx-reference-formatting & 0.00 & 0.25 & +0.25 \\
offer-letter-generator & 0.60 & 0.78 & +0.18 \\
react-performance-debugging & 0.50 & 0.67 & +0.17 \\
protein-expression-analysis & 0.00 & 0.17 & +0.17 \\
\bottomrule
\end{tabular}
\end{table}

\section{Method D Native Verifier Results}
\label{app:method_d}

Table~\ref{tab:main} reports the mean continuous reward under SkillsBench's native verifiers using Method~D scoring~\cite{skillsbench2026}. Method~D computes the per-task mean reward across $k$ trials, then averages across all 77 tasks with a fixed denominator. These continuous-reward results corroborate the binary pass-rate findings in the main text: \method{} consistently outperforms both baselines, though the absolute numbers are lower because partial credit is averaged rather than thresholded.

\begin{table}[h]
\centering
\caption{Method~D results (77 tasks, $k{=}2$ trials). Mean reward is the task-level mean across $k$ trials, then averaged across all 77~tasks with a fixed denominator. 95\% CI from bootstrap over tasks.}
\label{tab:main}
\begin{tabular}{lccc}
\toprule
\textbf{Condition} & \textbf{Mean Reward (Method D)} & \textbf{95\% CI} & \textbf{Gap Closed} \\
\midrule
No-skill   & 0.249 & $\pm$0.085 & -- \\
LLM-self   & 0.258 & $\pm$0.088 & -- \\
\method{}  & 0.273 & $\pm$0.084 & 11.4\% \\
Human      & 0.460 & $\pm$0.098 & 100\% \\
\bottomrule
\end{tabular}
\end{table}

Consistent with Li et al.~\cite{skillsbench2026}, LLM-authored skills provide negligible improvement over bare agents under native verifiers: 25.8\% vs.\ 24.9\%. After one iteration of \method{}, mean reward rises to 27.3\%. The confidence intervals with $k{=}2$ are wide ($\pm$8--10pp); a third trial seed will tighten them.

\section{Execution Example: How Skills Prevent Crashes}
\label{app:exec_example}

The hierarchical decomposition (Table~\ref{tab:hierarchical}) shows that \method{}'s primary benefit is execution reliability. Figure~\ref{fig:exec_example} illustrates the mechanism with a concrete example: \textbf{offer-letter-generator}, a task where the no-skill agent crashes but the \method{}-improved skill enables successful completion.

\begin{figure}[h]
\centering
\small
\begin{minipage}[t]{0.47\textwidth}
\raggedright
\textbf{No-skill agent (FAILS)}\\[3pt]
\hrule\vspace{4pt}
The agent must fill a Word template with employee data from JSON, replacing placeholders like \texttt{\{\{CANDIDATE\_NAME\}\}}.\\[3pt]
\textbf{Agent's approach:}
\begin{verbatim}
for run in para.runs:
  if '{{' + key + '}}' in run.text:
    run.text = run.text.replace(
      placeholder, str(value))
\end{verbatim}
\vspace{2pt}
\textbf{Result:} Placeholders remain unfilled. Word splits \texttt{\{\{POSITION\}\}} across multiple XML runs (e.g., \texttt{Run~1:~\{\{POSI}, \texttt{Run~2:~TION\}\}}). The per-run search never matches.
\end{minipage}%
\hfill
\begin{minipage}[t]{0.47\textwidth}
\raggedright
\textbf{\method{} skill agent (SUCCEEDS)}\\[3pt]
\hrule\vspace{4pt}
The \texttt{docx-template-fill} skill warns:\\[3pt]
\colorbox{green!15}{\parbox{0.92\linewidth}{\scriptsize\textbf{The \#1 issue with Word templates}: Word often splits placeholder text across multiple XML runs. \texttt{\{\{CANDIDATE\_NAME\}\}} might become Run~1: \texttt{\{\{CANDI}, Run~2: \texttt{DATE\_NAME\}\}}.}}\\[4pt]
\textbf{Skill's prescribed approach:}
\begin{verbatim}
full_text = paragraph.text
if placeholder in full_text:
  # rebuild with replacement
\end{verbatim}
\vspace{2pt}
\textbf{Result:} The agent joins runs at the paragraph level before matching, correctly replacing all placeholders.
\end{minipage}
\caption{Execution reliability example: \textbf{offer-letter-generator}. Without skills, the agent uses a naive per-run replacement that silently fails on split XML runs. The \method{}-improved skill prescribes paragraph-level replacement, preventing the crash.}
\label{fig:exec_example}
\end{figure}

This pattern recurs across the +26pp coverage gap. Tasks like \textbf{econ-detrending-correlation} and \textbf{citation-check} (both: no-skill crash, \method{} reward 1.0) similarly benefit from skill-provided library installation sequences and format-handling strategies.

\section{Trigger Analysis}
\label{app:trigger_analysis}

The SpreadsheetBench results depend on trigger-based routing, where skill selection quality is as important as skill content. We examine whether \method{}'s trigger metric produces measurably better activation descriptions.

\begin{table}[h]
\centering
\caption{Intrinsic trigger metrics. Coverage Breadth measures diversity of positive trigger phrases; Negative Specificity and Boundary Sharpness measure separation from negative phrases. Human skills rarely include negative phrases, zeroing two of three metrics.}
\label{tab:trigger}
\begin{tabular}{lccccc}
\toprule
\textbf{Condition} & \textbf{Skills} & \textbf{Cov.\ Breadth} & \textbf{Neg.\ Spec.} & \textbf{Bnd.\ Sharp.} & \textbf{Score} \\
\midrule
Human & 186 & 0.296 & 0.006 & 0.006 & 20.5 \\
LLM-self & 22 & 0.443 & 0.085 & 0.085 & 40.9 \\
\method{} & 216 & 0.410 & 0.338 & 0.238 & \textbf{65.7} \\
\bottomrule
\end{tabular}
\end{table}

Table~\ref{tab:trigger} reveals a pattern worth examining. LLM-generated descriptions produce more diverse positive triggers than human authors (Coverage Breadth 0.41--0.44 vs.\ 0.30), reflecting LLMs' tendency to enumerate scenarios comprehensively. The differentiator is negative specification: human skills almost never include exclusion clauses (Neg.\ Specificity 0.006), and unimproved LLM skills include few (0.085, with 19/22 skills having zero negatives). \method{}'s guardrail step adds targeted negative phrases, raising specificity to 0.338 and boundary sharpness to 0.238.

\paragraph{Activation discrimination.} To validate that intrinsic trigger metrics correlate with actual discriminative behavior, we embed all skill descriptions and task instructions and measure how well each description matches its target task vs.\ all others (Table~\ref{tab:discrimination}).

\begin{table}[h]
\centering
\caption{Activation discrimination (23 tasks, 190-skill pool). Margin = cosine similarity to correct task minus max similarity to any wrong task.}
\label{tab:discrimination}
\begin{tabular}{lcccc}
\toprule
\textbf{Condition} & \textbf{Mean Margin} & \textbf{Correct Sim.} & \textbf{Max Wrong Sim.} & \textbf{Rank=1} \\
\midrule
Human & 0.051 & 0.381 & 0.330 & 56.5\% \\
LLM-self & 0.031 & 0.359 & 0.327 & 56.5\% \\
\method{} & \textbf{0.148} & \textbf{0.509} & 0.361 & \textbf{73.9\%} \\
\bottomrule
\end{tabular}
\end{table}

Improved descriptions achieve a 3$\times$ wider discrimination margin than human skills. The improvement is asymmetric: correct-skill similarity increases by +0.13 over human (0.509 vs.\ 0.381), while max-wrong similarity increases only +0.03, meaning descriptions are specifically more aligned to their target tasks. A sign test across 23 tasks confirms statistical significance ($p < 0.02$, binomial test).

\section{Additional Trigger Embedding Visualization}
\label{app:trigger_viz}

\begin{figure}[h]
\centering
\includegraphics[width=\textwidth]{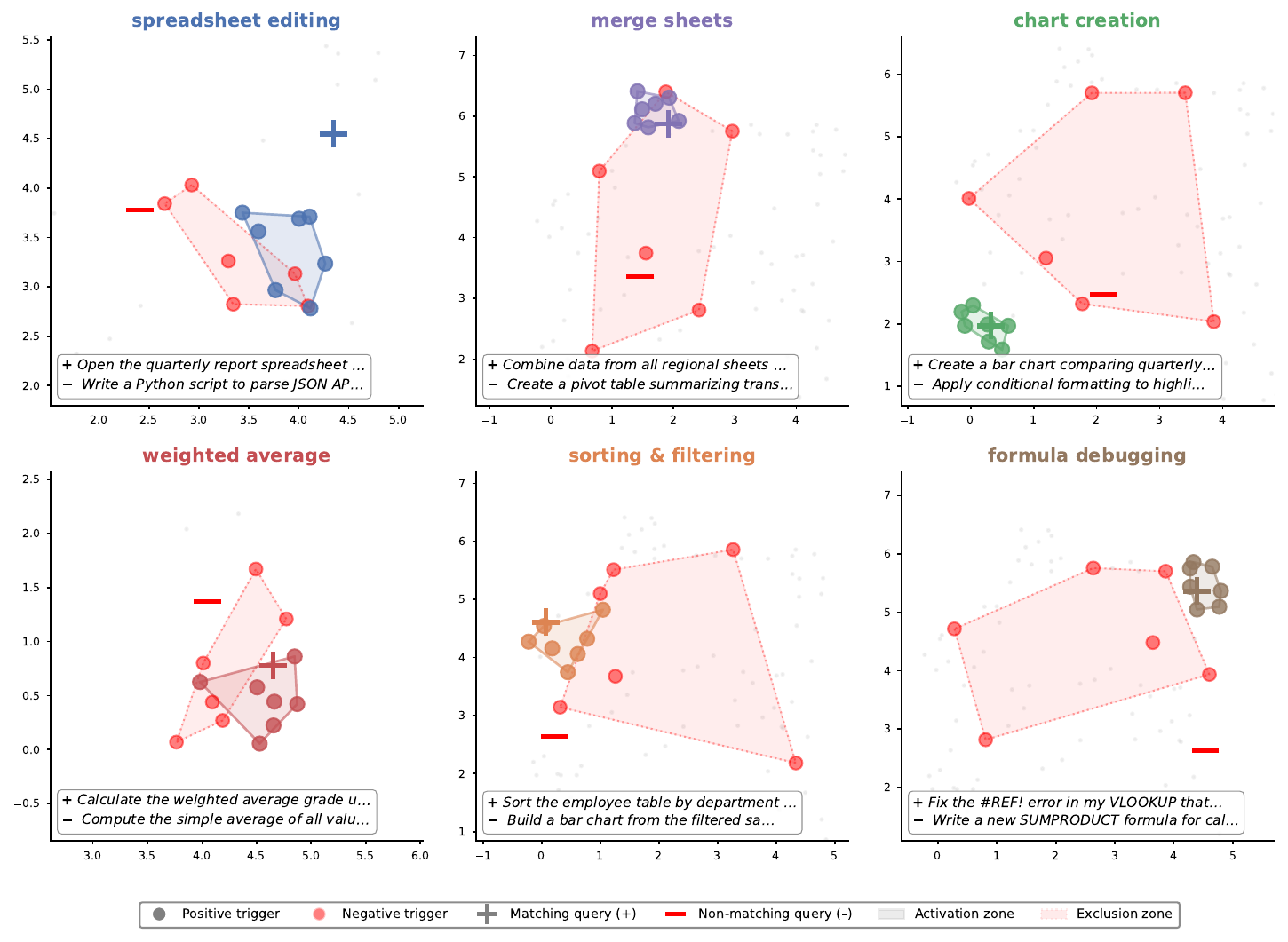}
\caption{UMAP projection of the trigger embedding space for six SpreadsheetBench library skills. Same visualization as Figure~\ref{fig:trigger_embedding} but within a single domain (spreadsheets). Even semantically related skills (\emph{merge sheets} vs.\ \emph{sorting \& filtering}) form distinct clusters with separated activation (\textbf{+}) and exclusion (\textbf{--}) zones.}
\label{fig:trigger_embedding_appendix}
\end{figure}

Figure~\ref{fig:trigger_embedding_appendix} shows the trigger embedding space for six skills from the SpreadsheetBench library, complementing the SkillsBench visualization in Figure~\ref{fig:trigger_embedding}. Despite all skills operating in the same spreadsheet domain, the trigger metric produces well-separated activation and exclusion zones.

\section{Instruction Compliance Rubric Details}
\label{app:rubric}

Section~\ref{sec:compliance} summarizes the two-call compliance evaluation. Here we provide the full scoring scales used by the LLM judge.

\paragraph{Call~1: Query-blind rule extraction.} Given only the skill document (without the task instruction), the judge extracts every actionable requirement and annotates each with severity:

\begin{table}[h]
\centering
\small
\begin{tabular}{@{}*{11}{p{4.4cm}}@{}}
\toprule
\textbf{Field} & \textbf{Values} & \textbf{Definition} \\
\midrule
\texttt{rule} & string & Concise statement of the requirement \\
\texttt{severity} & critical / major / minor & Core correctness rules (weight 3), important quality rules (weight 2), stylistic preferences (weight 1) \\
\bottomrule
\end{tabular}
\end{table}

\paragraph{Call~2: Query-aware per-rule judging.} Given the task instruction, the agent's output files (as cell-level diffs, rendered images, or text), and the extracted rules, the judge evaluates each rule on five dimensions. Tables~\ref{tab:rubric_score}--\ref{tab:rubric_fault} show the full scales.

\begin{table}[h]
\centering
\small
\caption{Adherence score scale ($a_i$): how well the agent followed the rule, judged from file evidence.}
\label{tab:rubric_score}
\begin{tabular}{@{}cl@{}}
\toprule
\textbf{Score} & \textbf{Meaning} \\
\midrule
1.0 & Clearly followed -- confirmed by file evidence \\
0.7 & Likely followed but evidence is incomplete \\
0.5 & Cannot determine -- insufficient evidence \\
0.3 & Likely violated \\
0.0 & Clearly violated -- evidence contradicts the rule \\
\bottomrule
\end{tabular}
\end{table}

\begin{table}[h]
\centering
\small
\caption{Rule quality scale ($g_i$): how specific and testable the rule is, independent of adherence.}
\label{tab:rubric_quality}
\begin{tabular}{@{}cp{9cm}@{}}
\toprule
\textbf{Quality} & \textbf{Meaning and examples} \\
\midrule
1.0 & Exact verifiable value (``blue text RGB: 0000FF'', ``format \$\#,\#\#0.00'') \\
0.8 & Specific constraint, verifiable (``zero formula errors'', ``use Excel formulas not hardcoded values'') \\
0.6 & Clear intent, partially verifiable (``use blue text for inputs'' -- blue, but which blue?) \\
0.4 & Directional guidance, hard to verify (``include clear section headers'', ``organize data logically'') \\
0.2 & Subjective or aesthetic (``professional appearance'', ``appropriate font'', ``clean layout'') \\
\bottomrule
\end{tabular}
\end{table}

\begin{table}[h]
\centering
\small
\caption{Skill-fault scale ($f_i$): when adherence is below 1.0, what fraction of the failure is attributable to the skill definition vs.\ the agent's execution?}
\label{tab:rubric_fault}
\begin{tabular}{@{}cp{9.5cm}@{}}
\toprule
\textbf{Fault} & \textbf{Meaning} \\
\midrule
0.0 & Entirely the agent's fault -- rule is clear and reasonable, agent did not follow it \\
0.3 & Mostly the agent's fault -- rule is slightly narrow but the agent did not even try \\
0.5 & Shared fault -- rule is over-constrained, agent got close but missed the exact spec \\
0.8 & Mostly the skill's fault -- rule is ambiguous or unclear \\
1.0 & Entirely the skill's fault -- rule is so vague or contradictory that no reasonable agent could follow it \\
\bottomrule
\end{tabular}
\end{table}

The fifth field, \texttt{evidence}, is a 1--2 sentence textual justification grounded in file-level observations (e.g., ``Diff confirms \texttt{Font(Color: \{rgb: \#0000FF\})} on input cells''). The \texttt{relevant} field is boolean: \texttt{false} when the rule's subject is absent from the output (e.g., ``green text for cross-sheet links'' on a task with no cross-sheet links). Irrelevant rules are excluded from all scoring and from the improvement brief.





\section{SpreadsheetBench Details}
\label{app:ssb}

\paragraph{Data split and conditions.} We split the 912-task SpreadsheetBench corpus into 200 train and 200 test tasks (random, seed=42). Libraries are built exclusively from train tasks; all reported numbers use the held-out test set. Three conditions are compared: (1)~\method{} library with embedding-based 3-zone routing and two-phase improvement, (2)~LLM-self library with LLM activation checks and no improvement, and (3)~no-skill baseline.

\paragraph{Agent and routing.} SSB experiments use an Excel copilot agent powered by Claude Opus 4.5, running with OfficeJS-based Excel automation. Skill library construction (generation, routing, and improvement) also uses Claude Opus 4.5. At test time, skills are \emph{discoverable but optional}: the agent's system prompt includes a manifest of available skill names and descriptions, and the agent may call a \texttt{get\_skill\_md} tool to load any skill's full content before planning its approach. This mirrors production deployment where the agent autonomously decides whether a skill is relevant, rather than having skills force-injected into the prompt. The \method{} library uses 3-zone embedding routing during \emph{construction}: for each training task, positive and negative trigger phrases are embedded with all-MiniLM-L6-v2 and affinity is computed as $\text{pos} - \text{neg}$ (cosine similarity against the skill's polarity phrases). The LLM-self library uses an LLM activation check (``does this task match this skill?'') for its binary fork/skip decision during construction.

\paragraph{Library construction.} Library building proceeds in two phases. \textbf{Phase~1} (no agent runs required): each train task instruction is routed through the library. The routing decision falls into one of three zones: FORK (no match--create a new skill), IMPROVE (match found--improve with trigger and solution-path diagnostics), or SKIP (near-duplicate--no action). Each skill collects up to 5 training tasks for improvement (capped to prevent overfitting to narrow patterns), with batch improvement applied after routing completes. \textbf{Phase~2} (requires agent trajectories): train tasks are run through the agent with their assigned skills, and each skill receives up to 1 additional improvement using quality impact and instruction compliance metrics from the actual trajectory. The LLM-self library uses the same FORK prompt but skips both improvement phases.

\paragraph{Evaluation methodology.} SpreadsheetBench~\cite{spreadsheetbench2024} evaluates by comparing cell values in agent-produced workbooks against golden answer workbooks at specified answer positions. A known challenge is that openpyxl (the Python library used for comparison) reads \emph{cached} formula values; when agents write new formulas, no cached values exist, causing false negatives. We address this by recalculating all output workbooks through Excel COM automation before evaluation: each workbook is opened in Excel, a full recalculation is triggered, and the workbook is saved to populate cached values.

\paragraph{Results analysis.} On the common evaluation subset ($N=50$ test tasks where all three conditions completed), both skill libraries dramatically improve over the no-skill baseline (52.0\% vs.\ 16.0\%, +36.0pp). The no-skill agent passes only 8 of 50 tasks, confirming that complex spreadsheet manipulation is extremely challenging without procedural guidance. Both the LLM-self and \method{} libraries fix an identical 18 additional tasks, raising pass rate to 52.0\%. The \method{} library achieves this with 22 skills versus the LLM-self library's 69, a 68\% reduction. Activation rates differ substantially: agents load a \method{} skill in 35.8\% of queries versus 20.0\% for the LLM-self library, indicating that consolidated, improved descriptions are more recognizable. Among \method{} skills, \texttt{vba-cell-processing-automation} is activated most frequently, reflecting the prevalence of VBA-based tasks in SpreadsheetBench.

\subsection{Skill Routing Distribution}
\label{app:ssb_routing}

Table~\ref{tab:routing_dist} shows how training tasks are distributed across \method{} skills during library construction. The distribution exhibits a power-law pattern characteristic of real-world skill libraries: a few generic skills attract many tasks, while specialized skills serve narrow but important niches.

\begin{table}[h]
\centering
\caption{Training task routing distribution across \method{} skills. ``Improve'' = tasks used for skill improvement (capped at 5); ``Skip'' = tasks routed to this skill but skipped as near-duplicates. Six skills never received training tasks (no match during routing) and remain in their initial generated form.}
\label{tab:routing_dist}
\small
\begin{tabular}{lrrr}
\toprule
\textbf{Skill} & \textbf{Improve} & \textbf{Skip} & \textbf{Total} \\
\midrule
formula-transform-lookup & 43 & 1 & 44 \\
formula-macro-data-processing & 29 & 0 & 29 \\
cross-sheet-aggregation & 20 & 8 & 28 \\
pattern-match-cross-reference & 20 & 6 & 26 \\
vba-cell-processing-automation & 20 & 2 & 22 \\
dynamic-range-aggregation & 12 & 3 & 15 \\
lookup-formula-troubleshooting & 9 & 1 & 10 \\
vba-data-reshaping & 9 & 0 & 9 \\
sequential-visit-counter & 3 & 1 & 4 \\
merge-deduplicate-columns & 0 & 3 & 3 \\
conditional-formula-aggregation & 2 & 1 & 3 \\
dynamic-expanding-average & 0 & 2 & 2 \\
range-lookup-by-threshold & 1 & 1 & 2 \\
multi-condition-intersection-lookup & 1 & 0 & 1 \\
array-transform-extract & 0 & 1 & 1 \\
delimiter-first-element-extraction & 1 & 0 & 1 \\
\emph{6 skills with 0 routed tasks} & 0 & 0 & 0 \\
\bottomrule
\end{tabular}
\end{table}

The top 5 skills absorb 75\% of all routed tasks (149/200), acting as broad patterns covering formula manipulation, VBA automation, and cross-sheet operations. These ``hub'' skills benefit most from the improvement loop, as they accumulate diverse failure evidence across many tasks. In contrast, 6 skills received no training tasks during routing and remain in their initial generated form---these represent edge-case patterns (e.g., weekday classification, date assembly) that are too specialized to match training tasks but may still serve rare test queries.

\section{Fair Grader Protocol}
\label{app:fairgrader}

The fair grader processes each of the 688 runs independently. For each run, the grader:

\begin{enumerate}[leftmargin=*]
    \item Collects the agent's output files from the Harbor workspace.
    \item Renders visual artifacts: Excel workbooks are converted to sheet-level images, PDFs are rendered page by page, and charts are exported as images.
    \item Constructs a multimodal prompt containing the task instruction, input file descriptions, and rendered output artifacts.
    \item A GPT-5.4 judge evaluates whether the agent substantively completed the task, producing a binary completion judgment, a confidence score (0--1), and a brief reasoning trace.
\end{enumerate}

When only agent log extracts are available (no preserved output files), these are labeled as \emph{unverified intent} and the grader defaults to incomplete unless the extracts contain actual final answer values. This conservative treatment means the fair grader underestimates true performance for conditions with low file preservation rates.



\end{document}